\documentclass[11pt]{article}
\usepackage{epsfig}

\newcommand{\text}{\rm}

\begin{document}

\title{\textbf{Renormalizability of the local composite operator }$A_{\mu }^{2}$%
\textbf{\ in linear covariant gauges}}
\author{D. Dudal\thanks{%
Research Assistant of The Fund For Scientific Research-Flanders, Belgium.} \
and H. Verschelde\thanks{%
david.dudal@ugent.be, henri.verschelde@ugent.be} \\
{\small {\textit{Ghent University }}}\\
{\small {\textit{Department of Mathematical Physics and Astronomy,
Krijgslaan 281-S9, }}}\\
{\small {\textit{B-9000 Gent, Belgium}}} \and V.E.R. Lemes, M.S. Sarandy,
R.F. Sobreiro, and S.P. Sorella\thanks{%
vitor@dft.if.uerj.br, sarandy@dft.if.uerj.br, sobreiro@dft.if.uerj.br,
sorella@uerj.br} \\
{\small {\textit{UERJ - Universidade do Estado do Rio de Janeiro,}}} \\
{\small {\textit{\ Rua S\~{a}o Francisco Xavier 524, 20550-013 Maracan\~{a},
}}} {\small {\textit{Rio de Janeiro, Brazil}}} \and J.A. Gracey\thanks{%
jag@amtp.liv.ac.uk } \\
{\small {\textit{Theoretical Physics Division}}} \\
{\small {\textit{Department of Mathematical Sciences}}} \\
{\small {\textit{University of Liverpool }}} \\
{\small {\textit{P.O. Box 147, Liverpool, L69 3BX, United Kingdom}}}}
\maketitle

\vspace{-13cm} \hfill LTH--588 \vspace{13cm}


\begin{abstract}
The local composite operator $A_{\mu }^{2}$ is analysed within the algebraic
renormalization in Yang-Mills theories in linear covariant gauges. We
establish that it is multiplicatively renormalizable to all orders of
perturbation theory. Its anomalous dimension is computed to two-loops in the
$\overline{\mathit{MS}}$ scheme.
\end{abstract}

\newpage

\section{Introduction}

The possibility that gluons might acquire a mass through a
dynamical mass generation mechanism is receiving increasing
attention, both from the theoretical point of view as well as from
lattice simulations. Effective gluon masses have been reported in
a rather
large number of gauges \cite{Dudal:2003tc}. For instance, the relevance of the local operator $%
A_{\mu }^{a}A^{a\mu }$ for Yang-Mills theory in the Landau gauge has been
emphasized by several authors \cite{gz,pb}. That this operator has a special
meaning in the Landau gauge follows by observing that, due to the
transversality condition $\partial _{\mu }A^{a\mu }=0$, the integrated mass
dimension two operator $\int d^{4}xA_{\mu }^{a}A^{a\mu }$ is gauge
invariant. Remarkably, the operator $A_{\mu }^{a}A^{a\mu }$ in the Landau
gauge is multiplicatively renormalizable \cite{vkav,jg}, its anomalous
dimension being given \cite{jg,dvs} by a combination of the gauge beta
function, $\beta (a)$, and of the anomalous dimension, $\gamma _{A}(a)$, of
the gauge field, according to the relation
\begin{equation}
\gamma _{A^{2}}(a)~=~-~\left( \frac{\beta (a)}{a}~+~\gamma _{A}(a)\right)
,\,\,\,\,\,\,\,\,\,\,\,a=\frac{g^{2}}{16\pi ^{2}}\,\,\,.  \label{ada2}
\end{equation}
Moreover, lattice simulations \cite{pb} have provided strong indications of
the existence of the condensate $\left\langle A_{\mu }^{a}A^{a\mu
}\right\rangle $, which is deeply related to the dynamical gluon mass. A
renormalizable effective potential for this condensate in pure Yang-Mills
theory has been constructed and evaluated in analytic form up to two-loop
order in \cite{vkav}, resulting in an effective gluon mass $m_{\mathrm{gluon}%
}\approx 500MeV$. The inclusion of massless quarks has been
recently worked out \cite{Browne:2003uv}. Another analytic study
of $\left\langle A_\mu^a A^{a\mu}\right\rangle$ can be found in
\cite{Dudal:2003vv}. Also, lattice simulations \cite{lrg} of the
gluon
propagator in the Landau gauge have reported a gluon mass $m_{\mathrm{gluon}%
}\approx 600MeV$. Concerning other gauges, an effective gluon mass
has been reported in lattice simulations in the Laplacian
\cite{lg} and maximal abelian \cite{as,bc} gauges. It is worth
underlining that the local operator $A_{\mu }^{a}A^{a\mu }$ of the
Landau gauge can be generalized \cite{k} to the maximal abelian
gauge, which is a renormalizable gauge in the continuum
\cite{mag1,fz,mag2}. It turns out in fact that the integrated
mixed
gluon-ghost operator$\footnote{%
In the case of the maximal abelian gauge the group index $\alpha $ labels
the off-diagonal generators $T^{\alpha}$ of $SU(N)$, with $%
\alpha=1,...,N(N-1)$. The parameter $\xi$ is the gauge fixing parameter of
the maximal abelian gauge.}$ $\int d^{4}x\left( \frac{1}{2}A_{\mu }^{\alpha
}A^{\alpha \mu }+\xi \overline{c}^{\alpha }c^{\alpha }\right) $ is BRST\
invariant on-shell \cite{k}, a property which ensures the multiplicative
renormalizability to all orders of perturbation theory \cite{ew,nos} of the
local operator $\left( \frac{1}{2}A_{\mu }^{\alpha }A^{\alpha \mu }+\xi
\overline{c}^{\alpha }c^{\alpha }\right) $. The analytic evaluation of the
effective potential for the condensate $\left\langle \frac{1}{2}A_{\mu
}^{\alpha }A^{\alpha \mu }+\xi \overline{c}^{\alpha }c^{\alpha
}\right\rangle $ has not yet been worked out. Nevertheless, we expect a
nonvanishing value for this condensate, which would result in a dynamical
gluon mass. This is supported by the fact that a renormalizable effective
potential for the mixed gluon-ghost operator has been obtained \cite{nos1}
in the nonlinear Curci-Ferrari gauge, yielding a nonvanishing condensate $%
\left\langle \frac{1}{2}A_{\mu }^{2}+\xi \overline{c}c\right\rangle $, which
provides a dynamical mass for the gluons. The Curci-Ferrari gauge shares a
close similarity with the maximal abelian gauge. We expect thus that
something similar should happen in this gauge.

A gluon condensate $\left\langle A_{\mu }^{a}A^{a\mu
}\right\rangle $ has also been introduced in the Coulomb gauge
\cite{gr} in order to obtain estimates for the glueball spectrum.
Older works \cite{f1,f2,gu,cs} already discussed the pairing of
gluons in connection with a mass generation, as a result of the
instability of the perturbative Yang-Mills vacuum. Also, the
dynamical mass generation for the gluons is a part of the
Kugo-Ojima criterion for color confinement \cite{ko,ko1}. See
\cite{rpt} for a review.

In this work we analyse the ultraviolet properties of the local composite
operator $A_{\mu }^{a}A^{a\mu }$ in the linear covariant gauges, whose gauge
fixing term is
\begin{equation}
\int d^{4}x\left( b^{a}\partial _{\mu }A^{a\mu }+\frac{\alpha }{2}b^{a}b^{a}+%
\overline{c}^{a}\partial ^{\mu }D_{\mu }^{ab}c^{b}\;\right) \;,  \label{gf}
\end{equation}
where $b^{a}$ stands for the Lagrange multiplier and $\alpha $ is the gauge
parameter. Our aim is that of establishing some necessary requirements in
order to study the possible condensation of this operator, which would imply
the occurrence of  dynamical mass generation in these gauges. Notice that,
unlike the case of the Landau and maximal abelian gauges, the quantity $\int
d^{4}xA_{\mu }^{2}$ is now not BRST\ invariant on-shell. However, we shall
be able to prove that the local operator $A_{\mu }^{2}$ is multiplicatively
renormalizable to all orders of perturbation theory. There is a simple
understanding of this property. In linear covariant gauges, due to the
additional shift symmetry of the antighost, \textit{i.e.} $\overline{c}%
\rightarrow \overline{c}+\mathrm{const.}$, the operator $A_{\mu }^{2}$
doesn't mix with the other local dimension two composite ghost operator $%
\overline{c}c$, which cannot show up due to the above symmetry. We remark
that the renormalizability of $A_{\mu }^{2}$ is the first step towards the
construction of a renormalizable effective potential in order to study the
possible condensation of this operator and the ensuing dynamical mass
generation.

The work is organized as follows. In Sect.2 we derive the Ward identities
for Yang-Mills theory in linear covariant gauges in the presence of the
local operator $A_{\mu }^{2}$. These identities turn out to ensure the
multiplicative renormalizability of $A_{\mu }^{2}$. In Sect.3 the explicit
two-loop calculation of the anomalous dimension of $A_{\mu }^{2}$ is
presented.

\section{Algebraic proof of the renormalizability of the local operator $%
A_{\mu }^{a}A^{a\mu }$}

We begin by recalling the expression of the pure Yang-Mills action
in the linear covariant gauges
\begin{eqnarray}
S &=&S_{YM}+S_{GF+FP}  \label{sym} \\
&=&-\frac{1}{4}\int d^{4}xF_{\mu \nu }^{a}F^{a\mu \nu }+\int d^{4}x\left(
b^{a}\partial _{\mu }A^{a\mu }+\frac{\alpha }{2}b^{a}b^{a}+\overline{c}%
^{a}\partial ^{\mu }D_{\mu }^{ab}c^{b}\right) \;,  \nonumber
\end{eqnarray}
where
\begin{equation}
D_{\mu }^{ab}\equiv \partial _{\mu }\delta ^{ab}-gf^{abc}A_{\mu }^{c}\;.
\label{cd}
\end{equation}
In order to study the local composite operator $A_{\mu }^{a}A^{a\mu }$, we
introduce it in the action by means of a BRST\ doublet \cite{book} of
external sources $\left( J,\lambda \right) $, namely
\begin{equation}
S_{J}=s\int d^{4}x\left( \frac{1}{2}\lambda A_{\mu }^{a}A^{a\mu
}+\frac{\xi }{2}\lambda J\right) =\int d^{4}x\left(
\frac{1}{2}JA_{\mu }^{a}A^{a\mu }+\lambda A^{a\mu }\partial ^{\mu
}c^{a}+\frac{\xi }{2}J^{2}\right) \;, \label{sj}
\end{equation}
where $s$ denotes the BRST\ nilpotent operator acting as
\begin{eqnarray}
sA_{\mu }^{a} &=&-D_{\mu }^{ab}c^{b}\;,  \nonumber \\
sc^{a} &=&\frac{1}{2}gf^{abc}c^{b}c^{c}\;,  \nonumber \\
s\overline{c}^{a} &=&b^{a}\;,  \nonumber \\
sb^{a} &=&0\;,  \nonumber \\
s\lambda &=&J\;,  \nonumber \\
sJ &=&0\;.  \label{s}
\end{eqnarray}
According to the local composite operators technique \cite{v1}, the
dimensionless parameter $\xi $ is needed to account for the divergences
present in the vacuum Green function $\left\langle
A^{2}(x)A^{2}(y)\right\rangle $, which turn out to be proportional to $J^{2}$%
. As is apparent from expressions (\ref{sym}) and (\ref{sj}), the action $%
\left( S_{YM}+S_{GF+FP}+S_{J}\right) $ is BRST\ invariant
\begin{equation}
s\left( S_{YM}+S_{GF+FP}+S_{J}\right) =0\;.  \label{si}
\end{equation}

\subsection{Ward Identities}

In order to translate the BRST\ invariance (\ref{si}) into the corresponding
Slavnov-Taylor identity \cite{book}, we introduce two further external
sources $\Omega _{\mu }^{a}$, $L^{a}$ coupled to the non-linear BRST\
variations of $A_{\mu }^{a}$ and $c^{a}$

\begin{equation}
S_{\mathrm{ext}}=\int d^{4}x\left( -\Omega ^{a\mu }D_{\mu }^{ab}c^{b}+\frac{1%
}{2}gf^{abc}L^{a}{c}^{b}c^{c}\right) \;,  \label{sext}
\end{equation}
with
\begin{equation}
s\Omega _{\mu }^{a}=sL^{a}=0\;.  \label{slo}
\end{equation}
Therefore, the complete action
\begin{equation}
\Sigma =S_{YM}+S_{GF+FP}+S_{J}+S_{\mathrm{ext}}\;,  \label{ca}
\end{equation}
obeys the following identities

\begin{itemize}
\item  The Slavnov-Taylor identity
\begin{equation}
\mathcal{S}(\Sigma )=\int d^{4}x\left( \frac{\delta \Sigma }{\delta A_{\mu
}^{a}}\frac{\delta \Sigma }{\delta \Omega ^{a\mu }}+\frac{\delta \Sigma }{%
\delta c^{a}}\frac{\delta \Sigma }{\delta L^{a}}+b^{a}\frac{\delta \Sigma }{%
\delta \overline{c}^{a}}+J\frac{\delta \Sigma }{\delta \lambda
}\right) =0\;. \label{sti}
\end{equation}

\item  The linear gauge-fixing condition
\begin{equation}
\frac{\delta \Sigma }{\delta b^{a}}=\partial _{\mu }A^{a\mu }+\alpha b^{a}\;.
\label{gfb}
\end{equation}

\item  The antighost equation
\begin{equation}
\frac{\delta \Sigma }{\delta \overline{c}^{a}}+\partial _{\mu }\frac{\delta
\Sigma }{\delta \Omega ^{a\mu }}=0\;.  \label{ageq}
\end{equation}
$\;$
\end{itemize}

Notice also that the additional shift symmetry in the antighost present in
the linear covariant gauges
\begin{equation}
\overline{c}\rightarrow \overline{c}+\;\mathrm{const.}  \label{ag}
\end{equation}
is automatically encoded in the antighost equation $\left( \mathrm{{\ref
{ageq}}}\right) $. Indeed, integrating expression $\left( \mathrm{{\ref{ageq}%
}}\right) $ on space-time yields

\begin{equation}
\int d^{4}x\frac{\delta \Sigma }{\delta \overline{c}^{a}}=0\;,  \label{iageq}
\end{equation}
which expresses in a functional form the shift symmetry $\left( \mathrm{{\ref
{ag}}}\right) $. Equations $\left( \mathrm{{\ref{ageq}}}\right) $, $\left(
\mathrm{{\ref{iageq}}}\right) $ imply that the antighost field can enter
only through the combination $\left( \Omega ^{a\mu }+\partial ^{\mu }%
\overline{c}^{a}\right) $, forbidding the appearance of the counterterm $%
\overline{c}^{a}c^{a}$. As a consequence, the local operator $A_{\mu
}^{a}A^{a\mu }$  does not mix with $\overline{c}^{a}c^{a}$ in linear $\alpha $%
-gauges.

Let us also display, for further use, the quantum numbers of all fields and
sources entering the action $\Sigma $
\begin{equation}
\stackrel{}{
\begin{tabular}{|c|c|c|c|c|c|c|c|c|}
\hline
& $A_{\mu }$ & $c$ & $\overline{c}$ & $b$ & $\lambda $ & $J$ & $\Omega $ & $%
L $ \\ \hline
$\dim \mathrm{.}$ & $1$ & $0$ & $2$ & $2$ & $2$ & $2$ & $3$ & $4$ \\ \hline
$\mathrm{gh-number}$ & $0$ & $1$ & $-1$ & $0$ & $-1$ & $0$ & $-1$ & $-2$ \\
\hline
\end{tabular}
}  \label{fields-table}
\end{equation}

\subsection{Algebraic characterization of the general local invariant
counterterm}

In order to characterize the most general local invariant counterterm which
can be freely added to all orders of perturbation theory \cite{book}, we
perturb the classical action $\Sigma $ by adding an arbitrary integrated
local polynomial $\Sigma ^{\mathrm{count}}$ in the fields and external
sources of dimension bounded by four and with zero ghost number, and we
require that the perturbed action $(\Sigma +\varepsilon \Sigma ^{\mathrm{%
count}})$ satisfies the same Ward identities and constraints as $\Sigma $ to
the first order in the perturbation parameter $\varepsilon ,$ \textit{i.e.}
\begin{eqnarray}
\mathcal{S}(\Sigma +\varepsilon \Sigma ^{\mathrm{count}}) &=&0+O(\varepsilon
^{2})\;,  \nonumber \\
\frac{\delta (\Sigma +\varepsilon \Sigma ^{\mathrm{count}})}{\delta b^{a}}
&=&\partial ^{\mu }A_{\mu }^{a}+\alpha b^{a}\;+O(\varepsilon ^{2})\;,
\nonumber \\
\left( \frac{\delta }{\delta \overline{c}^{a}}+\partial _{\mu }\frac{\delta
}{\delta \Omega _{\mu }^{a}}\right) (\Sigma +\varepsilon \Sigma ^{\mathrm{%
count}}) &=&0\;+O(\varepsilon ^{2})\;.  \label{cont-1}
\end{eqnarray}
This amounts to impose the following conditions on $\Sigma ^{\mathrm{count}}$%
\begin{equation}
\mathcal{B}_{\Sigma }\Sigma ^{\mathrm{count}}=0\;,  \label{b1}
\end{equation}
and
\begin{equation}
\frac{\delta \Sigma ^{\mathrm{count}}}{\delta b^{a}}=0\;,  \label{b2}
\end{equation}
\begin{equation}
\frac{\delta \Sigma ^{\mathrm{count}}}{\delta \overline{c}^{a}}+\partial
_{\mu }\frac{\delta \Sigma ^{\mathrm{count}}}{\delta \Omega _{\mu }^{a}}=0\;,
\label{b3}
\end{equation}
where $\mathcal{B}_{\Sigma }$ is the nilpotent linearized operator
\begin{eqnarray}
&&\mathcal{B}_{\Sigma }=\int d^{4}x\left( \frac{\delta \Sigma }{\delta
A_{\mu }^{a}}\frac{\delta }{\delta \Omega ^{a\mu }}+\frac{\delta \Sigma }{%
\delta \Omega ^{a\mu }}\frac{\delta }{\delta A_{\mu }^{a}}+\frac{\delta
\Sigma }{\delta c^{a}}\frac{\delta }{\delta L^{a}}+\frac{\delta \Sigma }{%
\delta L^{a}}\frac{\delta }{\delta c^{a}}+b^{a}\frac{\delta }{\delta
\overline{c}^{a}}+J\frac{\delta }{\delta \lambda }\right) \;,  \nonumber \\
&&  \label{lb}
\end{eqnarray}
\begin{equation}
\mathcal{B}_{\Sigma }\mathcal{B}_{\Sigma }=0\;.  \label{lbn}
\end{equation}
Taking into account that $\left( J,\lambda \right) $ form a BRST\
doublet, from the general results on the cohomology of Yang-Mills
theories \cite{bbh} it turns out that the external sources $\left(
J,\lambda \right)$ can contribute only through terms which can be
expressed as pure $\mathcal{B}_{\Sigma }$-variations. It follows
thus that the invariant local counterterm $\Sigma
^{\mathrm{count}}$ can be parametrized as
\begin{equation}
\Sigma ^{\mathrm{count}}=-\frac{\sigma }{4}\int d^{4}xF_{\mu \nu
}^{a}F^{a\mu \nu }+\mathcal{B}_{\Sigma }\Delta ^{-1}\;,  \label{cnterm}
\end{equation}
where $\sigma $ is a free parameter and $\Delta ^{-1}$ is the most general
local polynomial with dimension $4$ and ghost number $-1$, given by
\begin{eqnarray}
\Delta ^{-1} &=&\int d^{4}x\left( a_{1}\Omega _{\mu }^{a}A^{a\mu
}+a_{2}L^{a}c^{a}+a_{3}\partial _{\mu }\overline{c}^{a}A^{a\mu }+\frac{a_{4}%
}{2}gf_{abc}\overline{c}^{a}\overline{c}^{b}c^{c}\right.   \nonumber \\
\;\;\;\; &&\;\;\;\;\;\;\left. +a_{5}b^{a}\overline{c}^{a}+a_{6}\frac{\lambda
}{2}A^{a\mu }A_{\mu }^{a}+a_{7}\alpha \lambda \overline{c}^{a}c^{a}+a_{8}%
\frac{\xi }{2}\lambda J\right) \;,  \nonumber \\
&&\;\;\;\;\;\;  \label{delta-1}
\end{eqnarray}
with $a_{1},.....,a_{8}$ arbitrary parameters. From the conditions (\ref{b2}%
), (\ref{b3}) it follows that
\begin{equation}
a_{3}=a_{1},\;\;\;a_{4}=a_{5}=a_{7}=0\;,  \label{az}
\end{equation}
so that $\Delta ^{-1}$ reduces to
\begin{equation}
\Delta ^{-1}=\int d^{4}x\left( a_{1}(\Omega _{\mu }^{a}+\partial _{\mu }%
\overline{c}^{a})A^{a\mu }+a_{2}L^{a}c^{a}+a_{6}\frac{\lambda
}{2}A^{a\mu }A_{\mu }^{a}+a_{8}\frac{\xi }{2}\lambda J\right) \;.
\label{deltafim}
\end{equation}
Notice that the vanishing of the coefficient $a_{7}$ implies the absence of
the counterterm $J\overline{c}^{a}c^{a}$. As already underlined, this
ensures that the operator $A^{a\mu }A_{\mu }^{a}$ does not mix with the
ghost operator $\overline{c}^{a}c^{a}$. Therefore, for the final form of the
invariant counterterm one obtains:
\begin{eqnarray}
\Sigma ^{\mathrm{count}} &=&\int d^{4}x\left( -\frac{(\sigma +4a_{1})}{4}%
F_{\mu \nu }^{a}F^{a\mu \nu }+a_{1}\partial _{\mu }A_{\nu }^{a}F^{a\mu \nu
}\right.   \nonumber \\
&&+a_{2}\Omega _{\mu }^{a}(D^{\mu }c)^{a}+a_{2}\partial _{\mu }\overline{c}%
^{a}(D^{\mu }c)^{a}+a_{1}\Omega _{\mu }^{a}(\partial ^{\mu }c)^{a}-a_{1}%
\overline{c}^{a}\partial ^{2}c^{a}  \nonumber \\
&&+\frac{1}{2}(2a_{1}+a_{6})J{A^{a}}_{\nu }{A}^{a\nu
}+(a_{1}+a_{6}-a_{2})\lambda \partial _{\mu }c^{a}{A}^{a\mu }  \nonumber \\
&&\left.
-\frac{a_{2}}{2}gf_{abc}L^{a}c^{b}c^{c}+\frac{a_{8}}{2}\xi
J^{2}\right) \;.  \label{contfim}
\end{eqnarray}
It remains to discuss the stability of the classical action \cite{book},
i.e. to check that $\Sigma ^{\mathrm{count}}$ can be reabsorbed in the
classical action $\Sigma $ by means of a multiplicative renormalization of
the coupling constant $g$, the parameters $\alpha $ and $\xi $, the fields $%
\left\{ \phi =A,c,\overline{c},b\right\} $ and the sources $\left\{ \Phi
=J,\lambda ,L,\Omega \right\} $, namely
\begin{equation}
\Sigma (g,\xi ,\alpha ,\phi ,\Phi )+\varepsilon \Sigma ^{\mathrm{count}%
}=\Sigma (g_{0},\xi _{0},\alpha _{0},\phi _{0},\Phi _{0})+O(\varepsilon
^{2})\;,  \label{stab}
\end{equation}
with the bare fields and parameters defined as
\begin{eqnarray}
A_{0\mu }^{a} &=&Z_{A}^{1/2}A_{\mu }^{a}\,\,\,,\,\Omega _{0\mu
}^{a}~=~Z_{\Omega }\Omega _{\mu }^{a}\,\,\,,\,\,\,g_{0}~=~Z_{g}g\;,
\nonumber \\
c_{0}^{a}
&=&Z_{c}^{1/2}c^{a}\,\,\,\,\,,\,\,\,L_{0}^{a}~=~Z_{L}L^{a}\,\,\,\,\,,\,\,\,%
\alpha _{0}~=~Z_{\alpha }\alpha \,\,\,,  \nonumber \\
\overline{c}_{0}^{a} &=&Z_{\overline{c}}^{1/2}\overline{c}%
^{a}\,\,\,\,\,,\,\,\,J_{0}~=~Z_{J}J\,\,\,\,\,\,\,\,\,,\,\,\xi _{0}~=~Z_{\xi
}\xi \,,  \nonumber \\
b_{0}^{a} &=&Z_{b}^{1/2}b^{a}\,\,\,\,,\,\,\,\,\lambda _{0}~=~Z_{\lambda
}\lambda \,\,\,\,\,\,.\,  \label{rec}
\end{eqnarray}
The parameters $\sigma $, $a_{1}$, $a_{2}$, $a_{6}$, $a_{8}$ turn out to be
related to the renormalization of the gauge coupling constant $g$, of $%
A_{\mu }^{a}$, $c^{a}$, $J$, $\lambda $, and $\xi $ , according to
\begin{eqnarray}
Z_{g} &=&1-\varepsilon \frac{\sigma}{2}\,\,\,\,,\,\,  \nonumber \\
Z_{A}^{1/2} &=&1+\varepsilon \left( \frac{\sigma}{2}+a_{1}\right) ,  \nonumber \\
Z_{c}^{1/2} &=&1-\varepsilon \left( \frac{a_{1}+a_{2}}{2}\right) \,\,,
\nonumber \\
Z_{J} &=&1+\varepsilon \left( a_{6}-\sigma \right) \;,  \nonumber \\
Z_{\lambda } &=&1+\varepsilon \left(
a_{6}+\frac{a_{1}-a_{2}-\sigma}{2}\right) \,,
\nonumber \\
Z_{\xi } &=&1+\varepsilon \left( a_{8}-2a_{6}+2\sigma\right) \,.  \nonumber \\
&&  \label{ren}
\end{eqnarray}
Concerning the other fields and the sources $\Omega _{\mu }^{a}$, $L^{a}$,
it can be verified that they are renormalized as
\begin{eqnarray}
Z_{\overline{c}}\,
&=&Z_{c}\,\,\,,\,\,\,\,\,\,\,\,\,\;\;\;\;\;\;\,\,\,\,\,\,%
\,Z_{b}~=~Z_{A}^{-1}\,\,\,\,,\,\,\,\,\,Z_{\Omega }~=~Z_{c}^{1/2}  \nonumber
\\
Z_{L} &=&Z_{A}^{1/2}\,\,,\,\,\,\,\,\,\,\,\,\,Z_{\alpha }~=~Z_{A}~.
\label{oqf}
\end{eqnarray}
This completes the proof of the multiplicative renormalizability of the
local composite operator $A_{\mu }^{2}$ in linear covariant gauges. Finally,
it is useful to observe that, from eqs.(\ref{ren}), one has
\begin{equation}
Z_{\lambda }~=Z_{J}~Z_{c}^{1/2}Z_{A}^{1/2},  \label{zj}
\end{equation}
from which it follows that the anomalous dimension of $A_{\mu }^{2}$ turns
out to be related to that of the composite operator $A_{\mu }^{a}\partial
^{\mu }c^{a}$%
\begin{equation}
\gamma _{A\partial c}~=~\gamma _{A^{2}}+\gamma _{c}+\gamma _{A}\;,
\label{gj}
\end{equation}
where $\gamma _{c}$, $\gamma _{A}$, $\gamma _{A^{2}}$, and $\gamma
_{A\partial c}$ are the anomalous dimensions of the Faddeev-Popov ghost $%
c^{a}$, of the gauge field $A_{\mu }^{a}$, of the operator $A_{\mu }^{2}$,
and of the composite operator $A_{\mu }^{a}\partial ^{\mu }c^{a}$, which are
defined as
\begin{eqnarray}
\gamma _{c} &=&\mu \partial _{\mu }\ln Z_{c}^{1/2}~,~~~~\gamma _{A}=\mu
\partial _{\mu }\ln Z_{A}^{1/2}~,~\;\;\gamma _{A^{2}}=\mu \partial _{\mu
}\ln Z_{J}~,~~~\gamma _{A\partial c}=\mu \partial _{\mu }\ln
Z_{\lambda }~~~,
\nonumber \\
&&  \label{ad}
\end{eqnarray}
where $\mu $ is the renormalization scale. As expected, property
(\ref{gj}) relies on the fact that $A_{\mu }^{a}\partial ^{\mu
}c^{a}$ is the BRST\ variation of $\frac{1}{2}A_{\mu }^{2}$,
\textit{i.e.}
\begin{equation}
s\frac{A^{a\mu }A_{\mu }^{a}}{2}=-A_{\mu }^{a}\partial ^{\mu }c^{a}\;.
\label{bv}
\end{equation}
Although we did not consider matter fields in the previous analysis, it can
be checked that the renormalizability of $A_{\mu }^{2}$ and the relation (%
\ref{gj}) remain unchanged if matter fields are included.

\section{Calculation of the two-loop anomalous dimension of $A_\mu^2$}

We now turn to the computation of the anomalous dimension of $A_{\mu }^{2}$
in an arbitrary linear gauge. The method exploits the lack of mixing in the
linear covariant gauges between $A_{\mu }^{2}$ and the other dimension two
Lorentz scalar zero ghost number operator $\bar{c}^{a}c^{a}$, which we have
already noted. For instance, in the Curci-Ferrari gauge although both
operators mix there is a combination, $\mathcal{O}$~$=$~$\mbox{\small{$%
\frac{1}{2}$}}A_{\mu }^{2}$~$+$~$\alpha \bar{c}^{a}c^{a}$, which remains
multiplicatively renormalizable. Prior to the proof of \cite{dvs} that the
anomalous dimension of $\mathcal{O}$ was related to the $\beta $-function
and the gluon anomalous dimension, $\gamma _{\mathcal{O}}(a)$ was explicitly
computed at three loops in $\overline{\mbox{MS}}$ in \cite{jg}. That method
involved substituting the operator in a ghost two-point function with a
non-zero momentum flowing through the operator itself and one external ghost
momentum nullified. This configuration allowed for the application of the
\textsc{Mincer} algorithm, \cite{mincer}, written in the symbolic
manipulation language \textsc{Form}, \cite{form,formmincer}. A ghost
two-point function was chosen to avoid the appearance of spurious infrared
infinities which would arise for this momentum configuration if the external
legs were gluons. To determine $\gamma _{A^{2}}(a)$ in the linear gauges we
are forced into the same approach as \cite{jg} due to the infrared issue
with gluon external legs. Hence, we have renormalized the momentum space
Green's function $\langle c^{a}(p)\mbox{\small{$\frac{1}{2}$}}[A_{\mu
}^{2}](-p)\bar{c}^{b}(0)\rangle $ where $p$ is the external momentum.
Clearly, this has no tree term and therefore to deduce $\gamma _{A^{2}}(a)$
at $n$-loops requires renormalizing the Green's function at $(n+1)$-loops as
the one loop term corresponds to the tree term of $\langle A_{\mu }^{a}(p)%
\mbox{\small{$\frac{1}{2}$}}[A_{\mu }^{2}](-p)A_{\nu
}^{b}(0)\rangle $. This is evident, for example, by drawing one
and two loop diagrams for the various Green's functions based on
the interactions of the Yang-Mills action, eq.(\ref{sym}). Since
the \textsc{Mincer} algorithm currently only extracts the simple
poles
in $\epsilon $ in dimensional regularization to three loops, where $d$~$=$~$%
4 $~$-$~$2\epsilon $, this means we have only computed $\gamma _{A^{2}}(a)$
to two loops. Though this will be sufficient to deduce the effective
potential of $A_{\mu }^{2}$ to one loop. The Feynman diagrams for our
Green's function are generated with \textsc{Qgraf}, \cite{qgraf}, and
converted into \textsc{Form} input notation, \cite{jg}. At one loop there is
one diagram which plays the role of the tree diagram and at two loops there
are $15$ diagrams. The bulk of the calculation, however, is in the
evaluation of the $314$ three loop graphs. Since there is no operator mixing
we can apply the rescaling technique of \cite{lv} for automatic multiloop
computations to find the renormalization constant $Z_{A^{2}}$. From this we
deduce
\begin{eqnarray}
\gamma _{A^{2}}(a) &=&\left[ \left( 35+3\alpha \right)
C_{A}-16T_{F}N_{\!f}\right] \frac{a}{6}  \nonumber \\
&&+~\left[ \left( 449+33\alpha +18\alpha ^{2}\right)
C_{A}^{2}-280C_{A}T_{F}N_{\!f}-192C_{F}T_{F}N_{\!f}\right] \frac{a^{2}}{24}%
~+~O(a^{3})  \label{a2}
\end{eqnarray}
in the $\overline{\mbox{MS}}$ scheme where $N_{\!f}$ is the number of quarks
and the colour group Casimirs are defined by $\mbox{Tr} \left( T^a T^b
\right)$~$=$~$T_F \delta^{ab}$, $T^a T^a$~$=$~$C_F I$ and $f^{acd} f^{bcd}$~$%
=$~$C_A \delta^{ab}$. In deriving (\ref{a2}) from the corresponding
renormalization constant we have verified that the double pole in $\epsilon$
is correctly reproduced for all $\alpha$. Moreover, (\ref{a2}) reduces to
the Landau gauge expression of \cite{rbjg}.

\section{Conclusion}

We have investigated the renormalizability of the dimension two operator $%
A_{\mu }^{2}$ in arbitrary covariant linear gauges in Yang-Mills theories,
due to the possibility that it might condense and develop a non-zero vacuum
expectation value. This would generalize to these gauges previous results
obtained in the Landau gauge \cite{gz,pb,vkav,Browne:2003uv}. One feature of
our analysis is that, unlike the Curci-Ferrari gauges \cite{jg,nos}, the
operator $A_{\mu }^{2}$ does not mix with the other dimension two local
composite operator $\overline{c}^{a}c^{a}$. This is a general feature of the
linear covariant $\alpha $-gauges, present also in the Landau gauge \cite
{dvs}, $\alpha =0$, which is a consequence of the additional shift symmetry
in the antighost (\ref{ag}). Importantly the operator $A_{\mu }^{2}$ can
thus be treated in isolation as it does not require any ghost dependent
operator.

Finally, we underline that the multiplicative renormalizability of $A_{\mu
}^{2}$ and the explicit knowledge of its anomalous dimension for all $\alpha
$, eq.(\ref{a2}), are central ingredients towards the construction of a
renormalizable effective potential for studying the possible condensation of
this operator and the related dynamical mass generation, as was carried out
in the Landau \cite{vkav,Browne:2003uv} and Curci-Ferrari \cite{nos1} gauges.

\section*{Acknowledgments}

The Conselho Nacional de Desenvolvimento Cient\'{i}fico e Tecnol\'{o}gico
(CNPq-Brazil), the SR2-UERJ and the Coordena{\c{c}}{\~{a}}o de Aperfei{\c{c}}%
oamento de Pessoal de N{\'\i}vel Superior (CAPES) are gratefully
acknowledged for financial support.

\end{document}